\documentstyle[12pt,epsf]{dubna98}
\begin{document}
\begin{center}
{\bf PION-PION INTERACTION AT LOW ENERGY\\}
\vspace*{1cm}
M.E. SAINIO\\
{\it Department of Physics, University of Helsinki\\}
{\it P.O. Box 9, FIN-00014 Helsinki, Finland}
\end{center}

\vspace*{.5cm}
\begin{abstract}
{\small A brief discussion of the elastic $\pi \pi$ scattering amplitude
to two loops in chiral perturbation theory is given. Some technical
details of the evaluation of two-loop diagrams are addressed, and results relevant for the 
pionium decay are presented.}
\end{abstract}

\vspace*{1cm}
\section {INTRODUCTION}

\noindent In this talk I'll briefly summarize the results of our two-loop calculation of the 
$\pi \pi$ amplitude\cite{bij1,bij2}. Also, some technical issues\cite{gs}, useful for 
the evaluation of two-loop graphs in a number of processes, are discussed.

The investigation of the $\pi \pi$ interaction is of particular interest at present, because
there are several experiments which will bring new information after a long break in the 
experimental activity. 
Of special interest in this workshop is the pionium experiment DIRAC (PS 212)
at CERN\cite{Dirac} which aims at measuring the lifetime of the $\pi^+ \pi^-$
atomic bound state.
The relevant quantity there is the difference of the I=0 and I=2 $s$-wave
scattering lengths, $a^0_0 - a^2_0$. The forthcoming K$_{l4}$ data from the KLOE detector 
starting to operate at DA$\Phi$NE in Frascati will also add substantially to our
understanding of the low-energy $\pi \pi$ interaction.
Similar data from BNL (E865) are already coming to analysis\cite{BNL}. An attempt
at IUCF to see pionium production in the reaction $p d \rightarrow ^3\! \!{\it He}\, \pi^+ \pi^-$
has given only the upper limit, 1 $pb$, for the production cross section\cite{IUCF}. The
standard method to measure $\pi \pi$ scattering at high energies is to study the pion
induced pion production, $\pi N \rightarrow \pi \pi N$. At low energies, however, the 
extraction of $\pi \pi$ information from this reaction is complicated by the fact that there
one-pion-exchange does not necessarily dominate.

From theoretical point of view the interest in the $\pi \pi$ interaction is well
motivated. The question is to what extent our picture of spontaneously
broken chiral symmetry is valid. The $\pi \pi$ studies will provide a test. The technique
employed here, chiral perturbation theory (ChPT), is a systematic approach to build the chiral
symmetry of QCD into the low-energy amplitudes. The tool to do that
is the effective Lagrangian, so the $\pi \pi$ studies provide also a testing ground for the
effective field theory techniques. Another aspect of particular
interest recently is the size of the quark-antiquark vacuum condensate\cite{st1,lw2}.
According to the standard picture the condensate should be ``large'', whereas the, so called, 
generalized ChPT allows for a ``small'' value for the condensate. Generally speaking, these two
values correspond respectively to a small, about 0.2, and a large, around 0.3, value for the 
isoscalar $s$-wave scattering length. 

Another technique to address the $\pi \pi$ problem
is also making progress, namely the lattice approach. The lattice method can give estimates
of both the vacuum condensate and the $\pi \pi$ scattering lengths. However, some questions
remain, especially the role of the dynamical quarks, or the relatively high pion
mass still involved. In any case, some encouraging results have already been 
obtained\cite{aoki}.

ChPT is a systematic expansion of the Green functions in terms
of external momenta and quark masses. A number of theoretical assumptions are invoked.
These include that the chiral symmetry of QCD is spontaneously broken, and that the low-energy
singularities of the theory are generated by the Goldstone degrees of freedom\cite{we1,lw1}.

In this talk the $\pi \pi$ interaction will be discussed in chiral $SU(2) \times SU(2)$.
A technical remark referring to the evaluation of some loop diagrams is included. 
Then a summary of the results relevant for the pionium system will be given together 
with some prospects for future work.

\section{EFFECTIVE LAGRANGIAN}

\noindent The chiral effective Lagrangian can be written as
\begin{eqnarray}
{\cal L}_{\rm eff} = {\cal L}_2 + {\cal L}_4 + {\cal L}_6 + ...
\end{eqnarray}
where the indices refer to the chiral counting, the number of derivatives of the
pion field or twice the power of the quark mass terms. The two-loop calculation involves 
the evaluation of the amplitude up to and including
terms ${\cal O}(p^6)$. As a consequence, the part ${\cal L}_6$ appears at tree level,
the part ${\cal L}_4$ at both tree- and one-loop level and finally the piece ${\cal L}_2$
at tree-, 1-loop and 2-loop level. It is convenient to adopt for ${\cal L}_2$ the form
\begin{eqnarray}
         {\cal L}_2 = \frac{F^2}{4} \left\langle D_\mu U \, D^\mu U^\dagger
 + U \chi^\dagger + U^\dagger \chi\right\rangle
\end{eqnarray}
where the brackets $\langle \, \rangle$ refer to trace over the flavour
indices and $F$ is the pion decay constant in the chiral limit. It is convenient to work with 
the field
\begin{eqnarray}
         U &=& \sigma + i \frac{\mbox{\boldmath$\phi$}}{F} \; , \; \; \; \;
\sigma^2 + \frac{\mbox{\boldmath$\phi$}^2}{F^2} = {\bf \mbox{1}} \; \; ,
         \nonumber \\
        \mbox{\boldmath$ \phi$} &=&
 \left( \mbox{$\begin{array}{cc} \pi^0
& \sqrt{2} \pi^+
         \\ \sqrt{2} \pi^- &- \pi^0 \end{array}$} \right) =
\phi^i \tau^i  \; , 
         \nonumber \\
         \chi &=& 2 B(\hat{m}{\bf \mbox{1}}+ i p), \; \; 
\hat{m}=\frac{1}{2}(m_u+m_d)
\end{eqnarray}
where $p$ is an external pseudoscalar source.
The second constant appearing in ${\cal L}_2$, $B$,  is proportional to the vacuum condensate.
The next order Lagrangian is of the type
\begin{eqnarray}
         {\cal L}_4 &=& \sum^{4}_{i=1} l_iP_i + \dots
\end{eqnarray}
where the $l_i$'s are the low-energy constants required by the one-loop renormalization
of the $\pi \pi$ amplitude.
The explicit form of the terms $P_i$ can be found e.g. in ref.\cite{gl2}. The form of
${\cal L}_6$ has been given in ref.\cite{fs}.

\section{TWO-LOOP INTEGRALS}

\noindent A number of processes contain the same topologies in two-loop graphs:
\begin{itemize}
\item vector and axialvector 2-point functions
\item $\pi \rightarrow e \nu \gamma$
\item scalar and vector form factors of the pion
\item $\pi \pi \rightarrow \pi \pi$
\item $\gamma \gamma \rightarrow \pi \pi$
\end{itemize}
As an example let us consider
the fish graph in figure 1 contributing to $\pi \pi$ scattering.
\begin{figure}[t]
\begin{center}
\mbox{\epsfysize=4cm \epsfbox{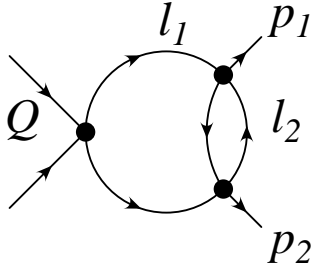}
     }
\caption{The fish diagram. The filled circles denote vertices from
the effective Lagrangian ${\cal L}_2$. The internal lines stand for scalar
propagators with mass 1.}
\label{fig1} \end{center} \end{figure}
The contribution is proportional to
\begin{eqnarray}
\int\frac{d^dl_1}{i(2\pi)^d} 
\int\frac{d^dl_2}{i(2\pi)^d} \, \prod_{i=1}^4\frac{1}{D_i}\nonumber
\end{eqnarray}
where
\begin{eqnarray}
D_1&=&1-l_1^2,\nonumber \\
D_2&=&1-(Q-l_1)^2, \nonumber \\
D_3&=&1-l_2^2,\nonumber \\
D_4&=&1-(l_2+l_1-p_1)^2, \nonumber \\
Q&=&p_1+p_2, \; \; Q^2=s. \nonumber
\end{eqnarray}
The integration over $l_2$ gives the loop--function $J(\bar{t})$
with $\bar{t}=(p_1-l_1)^2, \; \; \;$ ($w=d/2 - 2$)
\begin{eqnarray}
J(\bar{t})=C(w){\Gamma(-w)}\int_0^1dx\,[1-\bar{t}x(1-x)]^w,
\end{eqnarray}
which can be represented in the form
\begin{eqnarray}
J(\bar{t})=\int_4^\infty\frac{[d\sigma]}{\sigma-\bar{t}}.
\end{eqnarray}
The measure is
\begin{eqnarray}
[d\sigma]=\frac{C(w)\Gamma(3/2)}
{\Gamma(3/2 + w)}
 (\frac{\sigma}{4}-1)^w\beta \;d\sigma, \; \;{\rm where}\nonumber\\
 C(w)=\frac{1}{(4\pi)^{2+w}}, \; \; \;
\beta=(1-4/\sigma)^{1/2}.\nonumber
\end{eqnarray}
In the limit $d \rightarrow 4$ this gives
\begin{eqnarray}
\lim_{w\rightarrow
0}\;[d\sigma]=\frac{\beta}{16\pi^2}\;d\sigma.\nonumber
\end{eqnarray}
The emerging subdivergence is subtracted
$J(\bar{t})=J(0)+\bar{J}(\bar{t})$ and we have for the fish
diagram $J(s)J(0)+V(s)$ where
\begin{eqnarray}
V=\int_4^\infty
\frac{[d\sigma]}{\sigma} 
\int\frac{d^dl_1}{i(2\pi)^d}\,
\frac{\bar{t}}{D_1D_2
(\sigma-\bar{t})}\,.
\end{eqnarray}
Using the Feynman parametrization and integrating over $l_1$
yields in the limit $w \rightarrow 0$ a triple integral for the finite part, 
$V_f$. The imaginary part of $V_f$ can be constructed using the standard
techniques and the result
\begin{eqnarray}
{\rm Im}\, V_f(s) = \frac{\pi}{(16 \pi^2)^2} \left[ 3 \rho + \ln \frac{1-\rho}{1+\rho}-
\frac{1}{s \rho} \ln^2 \frac{1-\rho}{1+\rho} \right] \nonumber
\end{eqnarray}
follows. A function with the proper cut structure and $s \rightarrow
0,\infty$ behaviour is\cite{bij2,gs}
\begin{eqnarray}
V_f(s)=\frac{1}{(16 \pi^2)^2}
\left[\left( 3-\frac{\pi^2}{3 s \rho^2}\right) f +\frac{1}{2
\rho^2} f^2-
\frac{1}{3 s \rho^4} f^3  + 6 + \frac{\pi^2}{6} \right],
\end{eqnarray}
where
\begin{eqnarray}
f&=& \rho \left\{\ln\frac{1-\rho}{1+\rho} + i \pi
\right\}, \nonumber \\
\rho&=&\sqrt{1-4/s}, \; \; \;\; s > 4\;. \nonumber
\end{eqnarray}
The remarkable thing here is that even such graphs can be evaluated analytically.

\section{PION-PION AMPLITUDE}

\noindent The $\pi \pi$ amplitude can be written in the form
\begin{eqnarray}
T^{ik;lm} & = & \delta^{ik} \delta^{lm} A(s,t,u) + \delta^{il} \delta^{km}
A(t,s,u)  \nonumber \\
& + & \delta^{im} \delta^{kl} A(u,t,s) \nonumber
\end{eqnarray}
where to two loop order
\begin{eqnarray}
A(s,t,u)  =  \hspace{.3cm}{\it
x_2}\left[s-1\right] \hfill \; \; \; \; \; \; \; \; \; \; \; \;
\; \; \; \; \; \; \; \; \; \; \; \; \; \; \;\nonumber\\
+{\it x_2}^2\left[b_1+b _ 2s + b_3 s^2 +
b_4 ( t - u )^2\right]\nonumber\\
+{\it x_2}^2\left[F^{(1)}(s) +G^{(1)}(s,t)+G^{(1)}(s,u)\right]\nonumber\\
+{\it x_2}^3\left[b_5s^3+b_6s(t-u)^2\right]\nonumber\\
+{\it x_2}^3\left[F^{(2)}(s)+G^{(2)}(s,t)
+G^{(2)}(s,u)\right]\,\!\nonumber\\
+O({\it x_2}^4) 
\end{eqnarray}
and
\begin{eqnarray}
{\it x_2}=\frac{M_\pi^2}{F_\pi^2}.\nonumber
\end{eqnarray}
Unitarity fixes the analytically nontrivial
functions $F^{(1)}, G^{(1)}$ of the one-loop piece and $F^{(2)}, G^{(2)}$
for the two-loop amplitude\cite{bij1}.
The leading order result (${\cal O}(x_2)$) is due to Weinberg\cite{we2} and the
one-loop piece (${\cal O}(x_2^2)$) due to Gasser and Leutwyler\cite{gl1,gl2}.
In the isospin basis 
\begin{eqnarray}
T^0(s,t)&=&3 \, A(s,t,u) + A(t,u,s) + A(u,s,t) \nonumber \\
T^1(s,t)&=&A(t,u,s) - A(u,s,t) \nonumber \\
T^2(s,t)&=&A(t,u,s) + A(u,s,t) \nonumber
\end{eqnarray}
The relevant quantity for pionium decay is the
charge exchange amplitude
\begin{eqnarray}
T(\pi^+ \pi^- \rightarrow \pi^0 \pi^0) = 
\frac{1}{3}\left[T^0(s,t)-T^2(s,t)\right]\nonumber
\end{eqnarray}
and the decay rate
\begin{eqnarray}
{\rm RATE} \propto (a^0_0 - a^2_0)^2. \nonumber
\end{eqnarray}
The two-loop amplitude involves six low-energy constants (the one-loop level
has four $l$'s, and the two-loop piece six $r$'s\cite{bij1,bij2}. They appear,
however, in six independent combinations, $b_{1,...,6}$.) To clarify the structure
of these constants $b_4$ is taken as an example:
\begin{eqnarray}
b_4 = \frac{1}{2} \,l_2^r -\frac{1}{6}L
-\frac{1}{16{ \pi}^{2}}\frac{5}{36}\; \; \; \; \; \; \; \; \; \; \; \; \; \;
\; \; \; \; \; \; \; \; \; \nonumber \\
+ { x_2}
 \left\{ {\vrule height0.86em width0em depth0.86em} \right. \! \!
\frac{1}{16{ \pi}^{2}}\left[
{\vrule height0.86em width0em depth0.86em} \right. \! \!
\frac{10}{9}\,l_1^r + \frac{4}{9}\,l_2^r
-\frac{5}{9} \,l_4^r +\frac{47}{216} L \nonumber \\
 \; \; \; \; \; \; \; \; \; \; \; \;
+\frac{17}{3456}  +\frac{1}{16{ \pi}^{2}} \frac{1655}{2592}
\! \! \left. {\vrule height0.86em width0em depth0.86em} \right]
\nonumber \\
+ 2 \,l_2^r \,l_4^r -\frac{1}{6}\left[k_1+4 k_2 + k_4 \right] + r_4^r
\! \left. {\vrule height0.86em width0em depth0.86em}
 \right\}
\end{eqnarray}
where
\begin{eqnarray}
L&=&\frac{1}{16\pi^2}\ln{\frac{M_\pi^2}{\mu^2}} \nonumber \\
k_i&=&(4\,l_i^r(\mu)-\gamma_i L)L \; ;  \\
l_i^r&=&\frac{\gamma_i}{32\pi^2}\left(\bar{l}_i+
\ln\frac{M_\pi^2}{\mu^2}\right). \nonumber
\end{eqnarray}
In the framework of the generalized ChPT the $\pi \pi$ amplitude
has been evaluated\cite{kn1} in dispersive manner. The pieces fixed
by unitarity, $F$'s and $G$'s, agree with our results. The main difference
is that the Lagrangian approach taken here allows for determining the
structure of the constants $b_i$ of the polynomial part in terms of the
pion mass and the low-energy constants of both ${\cal O}(p^4)$ and ${\cal O}(p^6)$. 
The knowledge of the mass dependence makes it possible to compare results with
the lattice calculations where unphysical value for the pion mass is still
needed\cite{co}.

\section{RESULTS}

\begin{figure}[t]
\begin{center}
\mbox{\epsfysize=8cm \epsfbox{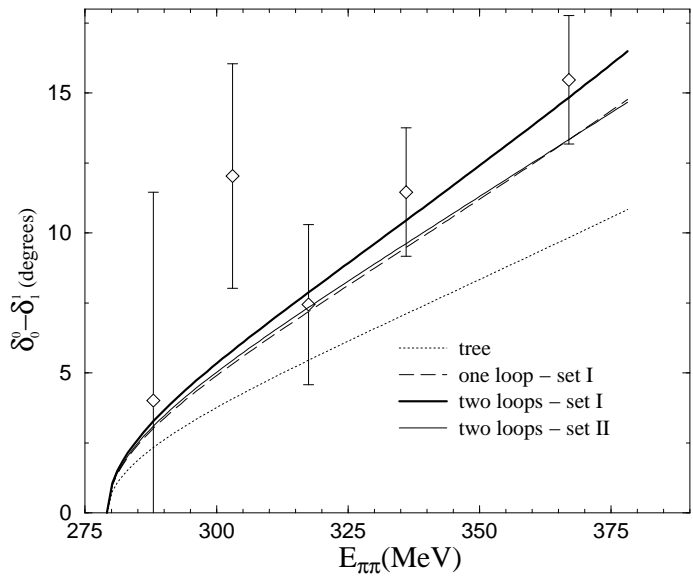}
     }
\caption{The phase shift difference $\delta^0_0-\delta^1_1$ for two different
sets of low-energy constants. The data points are from ref.\protect\cite{ro}.}
\label{fig2} \end{center} \end{figure}

\noindent For comparison with experiment numerical values for the constants 
$\bar{l}_{1,...,4}$ and $r$'s will be needed. The resonance saturation with the 
main contributions from the scalar and vector mesons can be used for the $r$'s. 
The influence of these values is small to the final amplitudes. 
The main effect comes from the constants $\bar{l}_i$ which show up already at 
order $x_2^2$. A detailed study of $\bar{l}_i$'s with constraints from the Roy
equations is on the way\cite{jg}, and with those values one can hope to reach
a precision of 2--3 \% for the $s$-wave scattering lengths. The values used
in ref.\cite{bij2} were taken from the K$_{l4}$ analysis of ref.\cite{bij3} or from
the $\pi \pi$ $d$-wave scattering lengths fixing $\bar{l}_1$ and $\bar{l}_2$, 
i.e. two sets of values were taken for the numerical comparisons. 
For the Set I we have\cite{bij3}:
\begin{eqnarray}
\bar{l}_1=-1.7, \; \; \; \bar{l}_2 = 6.1, \; \; \; \bar{l}_3=2.9, \; \; \; 
\bar{l}_4 = 4.3 \nonumber
\end{eqnarray}
and Set II:
\begin{eqnarray}
\bar{l}_1=-1.5, \; \; \; \bar{l}_2=4.5 \; \; \;
\bar{l}_3 \; {\rm and}\; \bar{l}_4 \; \mbox{{\rm as in SET I.}} \nonumber 
\end{eqnarray}
The phase shift difference $\delta^0_0-\delta^1_1$ for these sets of low-energy
constants has been given in figure \ref{fig2} as a function of the centre-of-mass
energy.

For the $s$-wave scattering length we get
\begin{eqnarray}
a^0_0 &=& \frac{7 {\it x_2}}{32 \pi } \left\{1 -{\it
x_2} \left[\frac{9}{2}L+\mbox{analytic}\right]\right.\nonumber\\
&&\left.\hspace{14.5mm}+{\it x_2}^2
\left[\frac{58}{7}k_1 +\frac{96}{7}k_2+5 k_3+\frac{11}{2} k_4
+\frac{1697}{84}\frac{L}{16\pi^2}+\mbox{analytic}\right]\right.\nonumber\\
&&\left.\hspace{9.5mm} +O({\it x_2}^3) \;\;\right\}
\end{eqnarray}
and numerically at scale $\mu$=1 GeV for the isoscalar $s$-wave scattering
length and the difference of the $s$-wave scattering lengths with Set I
\begin{eqnarray}
a_0^0&=& \overbrace{0.156}^{\mbox{tree}} +\overbrace{0.039 +
0.005}^{\mbox{1~loop}}+
\overbrace{0.013+0.003+0.001}^{\mbox{2~loops}}\;=\;
\overbrace{0.217}^{\mbox{total}}, \nonumber\\
 && \hspace{1.9cm} L \hspace{0.75cm}\mbox{anal.} \hspace{1.1cm} k_i
\hspace{1cm}L
\hspace{0.8cm}\mbox{anal.}\nonumber\\ \nonumber\\
a_0^0-a_0^2&=& 0.201 +0.036 +
0.006+
0.012+0.003+0.001\;=\;
0.258.\nonumber\\
 && \hspace{1.9cm} L \hspace{0.75cm}\mbox{anal.} \hspace{1.1cm} k_i
\hspace{1cm}L
\hspace{0.8cm}\mbox{anal.}\nonumber
\end{eqnarray}
In this case it is seen that the nonanalytic terms dominate both the one-loop
and the two-loop corrections.
The value for $a^0_0-a^2_0$ corresponding to Set II is 0.250. The experimental
figure obtained by using the universal curve is 0.29$\pm$0.04\cite{bij1}.
Using dispersion sum rules the constants $b_{3,...,6}$ have been determined\cite{kn2,gir}
and within the quoted errors these values are consistent with Set II.

\section{CONCLUSIONS AND FUTURE PROSPECTS}

\noindent The corrections in the loop expansion for the scattering length difference,
$a^0_0-a^2_0$, seem to be converging quite rapidly, about 20 \% correction from
the tree level to one-loop calculation and 6 \% from one-loop to two-loop
(calculated with Set I). Therefore,
in near future we can expect the precision of the calculation to reach 2-3 \% level
which is significant in strong interaction physics. This implies that the analyses
of $\pi \pi$ data will have to be made with extreme care. This is the challenge
for the K$_{e4}$ experiments in Brookhaven and Frascati, and the pionium
experiment at CERN.\\

\noindent{\bf Acknowledgements}\\

\noindent I would like to thank J\"urg Gasser for useful comments on the write-up.
Also, partial support from the EU-TMR programme, contract CT98-0169, is acknowledged.

\end{document}